\documentclass[10pt]{article}
\usepackage{latex8}
\usepackage{balance}

\font\tt=cmtt10 at 12pt
\ifx\pdfoutput\undefined
\usepackage{graphicx}
\usepackage[dvips]{epsfig}
\else
\usepackage[pdftex]{graphicx}
\fi
\usepackage{subfigure}
\usepackage{amsmath}
\usepackage{array}
\usepackage{latexsym}
{\bf}{\rm}
\begin{document}
%
\title{On the Requirements of New Software Development}

\author{Vincenzo De Florio and Chris Blondia \vspace*{3pt} \\
  University of Antwerp, Department of Mathematics and Computer Science\\
  Performance Analysis of Telecommunication Systems group\\
  Middelheimlaan 1, 2020 Antwerp, Belgium \vspace*{3pt} \\
  Interdisciplinary institute for BroadBand Technology\\
  Gaston Crommenlaan 8, 9050 Ghent-Ledeberg, Belgium}

\maketitle

%
\begin{abstract}
Changes, they use to say, are the only constant in life. Everything changes rapidly around us,
and more and more key to survival is the ability to rapidly adapt to changes. This consideration
applies to many aspects of our lives. Strangely enough, this nearly self-evident truth is not
always considered by software engineers with the seriousness that 
it calls for: The assumptions we draw for our systems
often do not take into due account that e.g., the run-time environments, the operational conditions,
or the available resources will vary. Software is especially vulnerable to this threat,
and with today's software-dominated systems controlling crucial services in
nuclear plants, airborne equipments, health care systems and so forth,
it becomes clear how this situation may potentially lead to catastrophes.
This paper discusses this problem and defines some of the requirements towards
its effective solution, which we call
``New Software Development'' as a software
equivalent of the well-known concept of New Product Development.
The paper also introduces and discusses a practical example of a software tool
designed taking those requirements into account---an adaptive data integrity provision
in which the degree of redundancy is not fixed once and for all
at design time, but rather it changes dynamically with respect to the disturbances experienced
during the run time.
\end{abstract}

\section{Introduction}\label{s:npd}
We live in a society of software-predominant systems. Computer systems are everywhere around us---from
supercomputers to tiny embedded systems---and we all know how important it is that services
appointed to computers be reliable, safe, secure, and flexible. What is often overlooked by many is the
fact that most of the logics behind those services, which support and sustain our societies, is in the
software layers. Software has become the point of accumulation of a large amount of complexity. Software
is ubiquitous, it is mobile, it has pervaded all aspects of our lives. Even more than that, the \emph{role\/}
appointed to software has become crucial, as it is ever more often deployed so as to fulfill mission
critical tasks. How was this possible in such a short time? The key ingredient to achieve this change in
complexity and role has been the conception of tools to manage the structure of software so as to divide
and conquer its complexity. 
\begin{itemize}
\item \emph{Dividing} complexity was achieved through specialization, by partitioning it
into system layers (for instance the OS, the middleware, the network, the application, and so forth).
\item \emph{Conquering\/} that complexity
was mainly reached by hiding it by means of clever organizations (for instance through object
orientation and, more recently, aspect and service orientation).
\end{itemize}
Unfortunately, though made transparent,
still this complexity is part of the overall system being developed. As a result, we have been given
tools to compose complex software-intensive systems in a relatively short amount of time, but the
resulting systems are often entities whose structure is unknown and are likely to be inefficient and
even error-prone. 

An example of this situation is given by the network software layers: We successfully decomposed the
complexity of our telecommunication services into well-defined layers, each of which is specialized on a
given sub-service, e.g. routing or logical link control. This worked out nicely when Internet was fixed.
Now that Internet is becoming predominantly mobile, those telecommunication services require complex
maintenance and prove to be inadequate and inefficient. Events such as network partitioning become the
rule, not the exception~\cite{CrFe99}. This means that the system and fault assumptions on which the original
telecommunication services had been designed, which were considered as permanently valid and hence
hidden and hardwired throughout the system layers, are not valid anymore in this new context. Retrieving
and exploiting this ``hidden intelligence"~\cite{eu08} is very difficult, which explains the many research
efforts being devoted world-wide to cross-layer optimization strategies and architectures.

Societal bodies such as enterprises or even governments have followed an evolutionary path similar to
that of software systems. Organizations such as enterprises or even governments have been enabled by
technology so as to deal with enormous amounts of data; still, like for software systems, they evolved
by trading ever increasing performance with ever more pronounced information hiding. The net result in
both cases is the same: Inefficiency and error-proneness. This is no surprise, as system-wide
information that would enable efficient use of resources and exploitation of economies of resources is
scattered into a number of separated entities with insufficient or no communication flow among each
other. This is true throughout the personnel hierarchy, up to the top: Even top managers nowadays only
focus on fragmented and limited ``information slices". Specialization (useful to partition complexity)
rules, as it allows an enterprise to become more complex and deal with a wider market. It allows
unprecedented market opportunities to be caught, hence it is considered as a panacea; but when the
enterprise is observed a little closer, often we observe deficiencies. In a sense, we often find out
that the enterprise looks like an aqueduct that for the time being serves successfully its purpose, but
loses most of its water due to leakage in its pipelines. This leakage is often leakage of structural
information---a hidden intelligence about an entity's intimate structure that once lost forbids any
``cross-layer" exploitation. Consequently efficiency goes down and the system (be it an enterprise, an
infrastructure, a municipality, or a state) becomes increasingly vulnerable: It risks to experience
failures or looses an important property with reference to competitiveness, i.e., agility,
that we define here as an entity's ability to reconfigure itself so as to maximize its ability
to survive and catch new opportunities. For business entities, a component of agility is the
ability to reduce time-to-market.
For software systems, this agility includes adaptability, maintainability, and  reconfigurability---that is,
adaptive fault-tolerance support. We are convinced that this property will be recognized in the
future as a key requirement for effective software development---the software equivalent of 
the business and engineering concept of New Product Development~\cite{NPD}.
The tool described in this paper---an adaptive data integrity provision---provides a practical example
of this vision of a ``New Software Development.''

The structure of this paper is as follows: 
Section~\ref{s:intro} introduces the problem of adaptive redundancy and data integrity.
Section~\ref{s:sota} is a brief discussion on the available data integrity provisions.
A description of our tool and design issues are given in Sect.~\ref{s:tool}.
In Sect.~\ref{s:perfanalysis} we report on an analysis of the performance of our tool.
Our conclusions are finally drawn in Sect.~\ref{s:end}.

\section{Adaptive Redundancy and Data Integrity}\label{s:intro}
A well-known result in information theory by Shannon~\cite{Shannon} tells us that, from any
channel with known and fixed unreliability, it is possible to set up a more reliable channel by
increasing the degree of information redundancy. 
A similar conclusion holds
in the domain of fault-tolerance: Let us consider a computer service that
needs to sustain a predefined level of
reliability when deployed in a given disturbed environment. Furthermore, let us
assume that such environment is characterized by a known and fixed scenario of disturbances---for instance,
known and bounded electro-magnetic interference.
In such a case, the required reliability can be reached by using some fault-tolerance mechanism
and a certain level of redundancy (in time, information, design, or hardware) to cope with
the disturbances. An example of this approach is described e.g. in~\cite{DeBo98}, where
a stable memory is obtained by using a spatial and temporal redundancy scheme.
In both cases, when the channel (respectively, the environment) does not change
their characteristics, a precise estimation of the unreliability (respectively,
the possible disturbances affecting the run-time environment) can effectively enhance
reliability.

Unfortunately, such a precise estimation is often not possible or not meaningful.
To show this, let us focus on a particular problem, namely that of 
data integrity: There, the goal
is being able to protect data against memory disruptions and other faults that could make it
impossible to access some previously saved data. Furthermore, let us consider a particular
solution to that problem, namely redundant data structures. In such case
redundancy and voting are used to protect memory from possible transient or permanent corruptions. A
common design choice in data integrity through redundant data structures is having
a static fault model assumption, such as e.g. ``during any
mission, up to 3 faults shall affect the replicas'', which translates into using a 7-redundant cell to
address the worst possible scenario. This brings us to a number of observations:

\begin{itemize}
\item First of all, such assumptions are the result of worst-case analyses which often are not
a viable solution, especially
in the case of embedded and mobile devices: A large consumption of resources 
would severely hamper the operating autonomy without bringing any tangible added value.

\item Moreover, a static fault model implies that
our data integrity provisions will have a fixed range of admissible events to address and
tolerate. This translates into two risks, namely

\begin{enumerate}
\item overshooting, i.e., over-dimensioning the data integrity provision with respect to the actual threat being experienced, and
\item undershooting, namely underestimating the threat in view of an economy of resources.
\end{enumerate}

Note how those two risks turn into a veritable dilemma to the designer: Wrong choices at this point
can lead to
either unpractical, too costly designs or cheap but vulnerable provisions.

\item 
Another tricky aspect in a static choice of the possible faults to affect our service
is the hidden dependence on the target hardware platform. This dependence translates 
in very different failure
semantics. In other words, the way the target hardware components behave in case of failure
can vary considerably with the adopted technology and even the particular component.
As an example, while yesterday's software was running atop CMOS chips,
today the common choice e.g. for airborne applications
is SDRAM---because of speed, cost, weight, power and simplicity of design~\cite{Lad02}. But CMOS memories
mostly experience single bit errors~\cite{1981coae.conf...66O}, while SDRAM chips are
known to be subjected to several classes of ``nasty'' faults, including 
so-called ``single-event effects''~\cite{Lad02}, i.e.,
single faults affecting whole chips. Examples include: 
\begin{enumerate}
\item Single-event latchup,
        a threat that can bring to the loss of all data stored on chip~\cite{WikiLatchup}.
\item Single-event upset (SEU), leading to frequent soft errors~\cite{WikiSoftError,WikiSEU}.
\item Single-event functional interrupt (SEFI), i.e. a special case of SEU that places the device
        into a test mode, halt, or undefined state. The SEFI halts normal operations, and requires
        a power reset to recover~\cite{Holbert}.
\end{enumerate}
Furthermore the already cited paper~\cite{Lad02} remarks how even from lot
to lot error and failure rates can vary more than one order of magnitude.
In other words, the superior performance of the new generation of memories is paid with a higher instability
and a trickier failure semantics.

Despite these veritable pitfalls, hardware dependence is mostly not taken into account when developing
fault-tolerant services e.g. for data integrity.
Ideally, hardware dependence and impact on failure semantics should be explicitly available
at service design, composition, and deployment times. Tools could then be conceived to highlight
risks and assist the designer of software services.

\item Still another aspect derives from the choice of the failure semantics:
Addressing a weak failure semantics, able to span
many failure behaviours, effectively translates
in higher reliability---nevertheless,
\begin{enumerate}
\item it \emph{requires\/} large amounts of extra resources,
      and therefore implies a high cost penalty, and
\item it \emph{consumes\/} those extra resources,
      which translates into their rapid exhaustion.
\end{enumerate}
For instance, a well-known result
by Lamport et al.~\cite{LaSP82} sets the minimum level of redundancy
required for tolerating Byzantine failures to a value that is
greater than the one required for tolerating, e.g., value failures.
Using the simplest of the algorithms described in the cited paper,
a 4-modular-redundant (4-MR) system can only withstand
any \emph{single Byzantine failure}, while the same system
may exploit its redundancy to withstand up to
three crash faults---though no other kind of
fault~\cite{Pow97a}. In other words:
\begin{quote}
After the occurrence of a crash fault,
a 4-MR system with strict Byzantine failure semantics has exhausted its
redundancy and is no more dependable than a non-redundant system supplying
the same service, while the crash failure semantics system is able to survive
to the occurrence of that and two other crash faults. On the other hand, the
latter system, subject to just one Byzantine fault, would fail regardless
its redundancy.
\end{quote}
We conclude that, for any given level of redundancy,
\emph{trading complexity of failure mode against number and type
of faults tolerated\/} may be considered as an important capability
for an effective fault-tolerant structure. Dynamic adaptability
to different environmental conditions\footnote{%
           The following quote by J. Horning~\cite{Hor98} captures
           very well how relevant may be the role of the environment
           with respect to achieving the required quality of service:
           ``What is the most often overlooked risk in software engineering?
           That the environment will do something the designer never
           anticipated.''}
can provide a satisfactory answer
to this need, especially when the additional complexity required to
manage this feature does not burden or jeopardize the application.

\item A final argument is that run-time environments change, often because services are
usually mobile but
sometimes also because of external events affecting e.g. temperature, radiation, electro-magnetic
interference, or cosmic rays. Applications such as a wireless sensor network to assist a fire brigade, 
or a spaceborne application orbiting around the sun, are \emph{bistable\/} by nature: They usually
operate in either a ``normal'' state, where the environmental conditions are not particuarly demanding,
or in a ``critical'' state, likely to affect the mission's integrity if no special countermeasures are
taken. A sudden fire or a solar flare are examples of events requiring to switch to critical state.
Also in such cases a static choice would be
unpractical, as the nature of faults is meant to change during a same mission.
\end{itemize}

Our vision to this problem is that fault-tolerant services
should be built with architectural and/or structuring
techniques able to decompose their complexity, but without hiding the basic hypotheses and
assumptions about their target execution environment, the expected fault- and system models, and
the failure semantics assumed for the components our services depend on.
Our conjecture is that effective solutions to this problem should come by \emph{considering the nature
of faults as a dynamic system}, i.e., a system evolving in time, and by expressing
the fault model as a function $F(t)$. Consequently, we believe that any fault-tolerance provision that 
be able to solve the above mentioned flaws 
should make use of an adaptative feedback loop~\cite{PVR06}. In such loop redundancy would be
allocated according to the measured values of $F(t)$, obtained by monitoring a set of
meaningful environmental variables. 

This paper reports on the design of a tool compliant to such model. 
Our tool allows designers to make use of adaptively redundant data structures
with commodity programming languages such as C or Java. Designers using our tool can define
redundant data structures where the degree of redundancy is not chosen at design time but
changes dynamically with respect to the disturbances experienced
at run-time. In other words, our approach attunes the
degree of redundancy required to ensure data integrity to the actual faults
being experienced by the system and provides an example of adaptive
fault-tolerance software provision. Such property allows to
reach adaptability, maintainability, and  reconfigurability, which we deem
as being three fundamental components to fulfill the requirements of ``new software
development''---the software equivalent of
the business and engineering concept of New Product Development~\cite{NPD}.

\section{A Brief Introduction to Data Integrity Provisions}\label{s:sota}
As well known, redundancy is key to error detection, 
correction, and recovery~\cite{TaMB80}. The
use of redundancy to detect and correct errors in 
stored data structures has been investigated
for decades. Techniques may be divided into 
two classes: Those that only support error detection and those that
support error correction as well.

Protection regions through codewording~\cite{BRSSS03} 
and structure marking~\cite{Strmark}
are examples of error detection techniques. 
The rationale of such methods is that,
when a ``wild store'' occurs, due to
a design or physical fault, or because of a malicious attack,
the techniques allow detecting the change\footnote{Following~\cite{TaMB80}, we
    define a \emph{change\/} as ``an elementary modification
    to the encoded form of a data structure instance.''}
and shutting down the service
so as to enforce crash failure semantics. This is in general an effective
approach, though there exist cases, e.g. unmanned spaceborn missions,
where other approaches could be more sensible. One such approach is
discussed in the rest of this paper. 

Error correction in general requires more resources
in time and space and ranges from corruption recovery in 
transaction processing systems through redoing~\cite{BRSSS03} to
full backup copies. Backup copies can be arranged through bitwise replicas (semantical
integrity) or by using redundancy in the representation of the 
data (structural integrity). Examples of the latter can be found
in~\cite{TaMB80}. An example of the former case is the method described
in our paper.

\section{Dynamically Redundant Data Structures}\label{s:tool}
\begin{figure}
\centerline{\includegraphics[width=0.45\textwidth]{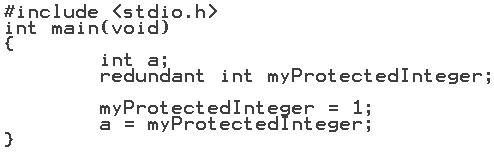}}
\caption{A simple example of use of redundant variables.}\label{f:ex1}
\end{figure}

\begin{figure}[t]
\centerline{\includegraphics[width=0.65\textwidth]{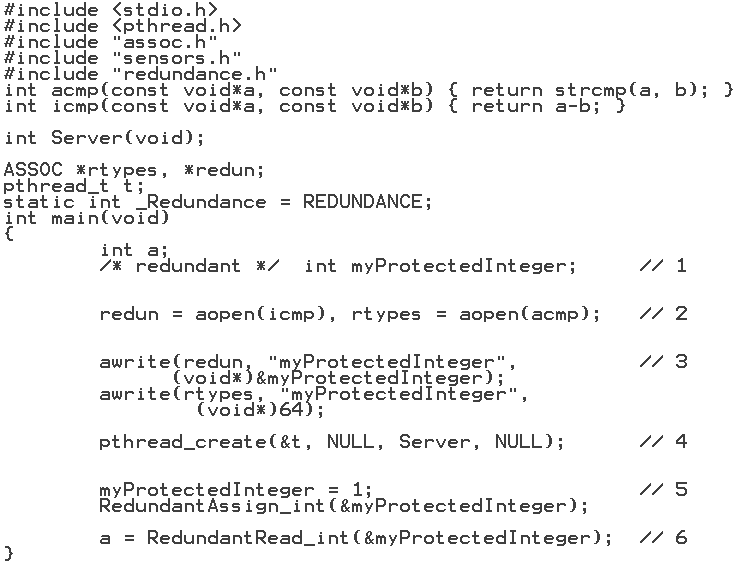}}
\caption{An excerpt from the translation of the code in Fig.~\ref{f:ex1}. Variable ``\_Redundance''
represents the current amount of redundancy, initially set to ``REDUNDANCE'' (that is, 3).}\label{f:tr1}
\end{figure}

Our tool is a translator that loosely parses a source code performing some transformations as reported
in the rest of this section. We developed our translator in the C language and with
Lex \& YACC~\cite{LeMB92}.
The reported version supports the C syntax though the same principles can be easily applied
to any other language. Our translator performs a simple task---it allows the programmer to tag
scalar and vector variables with a keyword, ``redundant,'' and then 
instruments the memory accesses to tagged variables.
Figure~\ref{f:ex1} shows how this is done in practice with a very simple example whose translation
is provided in Fig.~\ref{f:tr1}.
Let us
review the resulting code in more detail (please note that item x in the following list refer to lines
tagged as ``// {\em x}" in the code):

\begin{enumerate}
 \item First the translator removes the occurrences of attribute ``redundant".
 \item Then it performs a few calls to
     function ``aopen"~\cite{Dev27}. This is to open the associative arrays ``redun" and ``rtype".
     As well known, an associative array generalizes the concept of array so as
     to allow addressing items by non-integer indexes. The arguments to ``aopen"
     are functions similar to ``strcmp", from the C standard library, which are
     used to compare index objects. The idea is that these data structures
     create links between the name of variables and some useful information (see
     below).
 \item There follow a number of
     ``awrites", i.e., we create associations between variables' identifiers and
     two numbers: the corresponding variables' address and an internal code
     representing its type and attributes (code 64 means ``redundant int'').
 \item Then the ``Server" thread,
     responsible to allocate replicas and to monitor and adapt to external changes, is spawned.
 \item A write access to
     redundant variable {\em w}, of type {\em t}, is followed by a call to
     ``RedundantAssign\_{\em t}(\&{\em w})".
 \item Finally, reading from redundant
     variable {\em w}, of type {\em t}, is translated into a call to
     function ``RedundantRead\_{\em t}(\&{\em w})".
\end{enumerate}

The strategy to allocate replicas is a research topic on its own. Solutions range from na\"ive
simple strategies like allocating replicas into contiguous cells---which makes them
vulnerable to burst faults---to more sophisticated strategies where replicas get allocated
e.g. in different memory banks, or different memory chips, or even on different processors.
Clearly each choice represents a trade-off between robustness and performance penalty.
In our current version we separate replicas by strides of variable length.

The core of our tool is given by functions ``RedundantAssign\_{\em t}(\&{\em w})" and
``RedundantRead\_{\em t}(\&w)", which are automatically generated for each type {\em t} through
a template-like approach. The former function performs a redundant write, the latter a
redundant read plus majority voting. For voting, an approach similar to that
in~\cite{DeDL98e} is followed.

What makes our tool different from classical libraries for redundant data structures such as
the one in~\cite{TaMB80} is the fact that in our system the amount of replicas
of our data structures changes dynamically with respect to the observed disturbances. We assume
that a monitoring tool is available to assess the probability of memory corruptions of the
current environment. We also provide an example of such a monitoring tool, which estimates
that probability by measuring for each call to ``RedundantRead\_{\em t}'' the risk of failure $r$.
Quantity $r$ may be defined for instance as follows: If our current redundancy is $2n+1$, and if the maximum
set of agreeing replicas after a ``RedundantRead\_{\em t}'' is $m$, ($1\le m\le 2n+1$), then

\begin{equation}
r = \left\{ \begin{array}{ll}
(2n+1-m)/n & \quad\textrm{if $m>n$}\\
1& \quad\textrm{otherwise.}
\end{array} \right.
\label{eq:1}
\end{equation}

For instance if redundancy is 7 and $m=6$, that is if
only one replica differs, then $r=1/3$. Clearly the above choice of $r$ lets risk 
increase linearly with the number of replicas not in agreement with the majority. Other formulations
for $r$ and for the monitoring tool\footnote{It is clear that a proactive approach would be more
	effective than the one reported here, which requires to perform voting to assess
	the need for adaption.} are possible and likely to be more effective than the 
ones taken here---also a matter for future research.

Our strategy to adjust redundancy is also quite simple: If we observe that $r>0.5$, redundancy is
increased by 2; if $r=0$ for 1000 consecutive calls to ``RedundantRead\_{\em t}'', redundancy is
decreased by 2. Lower bound and upper bound for redundancy have been set to 3 and 11 respectively.

Each request for changing redundancy is reflected by the ``Server'' thread into variable
``\_Redundance'' through the scheme introduced in what follows.

\subsection{Reflective and Refractive Variables}
The idea behind reflective and refractive variables (RR vars)~\cite{DB07a}
is to use memory access as an abstraction to perform concealed tasks. RR vars
are volatile variables whose identifier links them to an external device, such as a sensor, or an
RFID, or an actuator. In reflective variables, memory cells get asynchronously updated by service
threads that interface those external devices. We use the well-known term ``reflection'' because those
variables in a sense ``reflect'' the values measured by those devices. In refractive 
variables, on the contrary, write
accesses trigger a request to update an external parameter, such as the data rate of the local TCP
protocol entity or the amount of redundancy to be used in transmissions. We use to say that write
accesses ``refract'' (that is, get redirected~\cite{TST}) onto corresponding external devices.

\begin{figure}
\centerline{\includegraphics[width=0.5\textwidth]{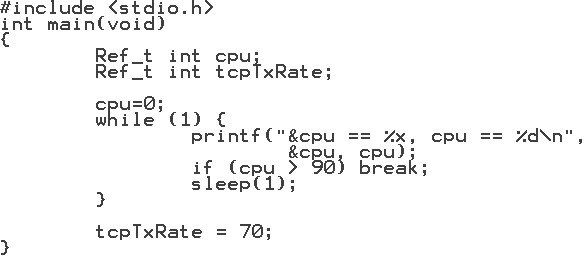}}
\caption{A simple example of the use of RR vars.}
\label{f:rr1}
\end{figure}

\begin{figure}[t]
\centerline{\includegraphics[width=0.75\textwidth]{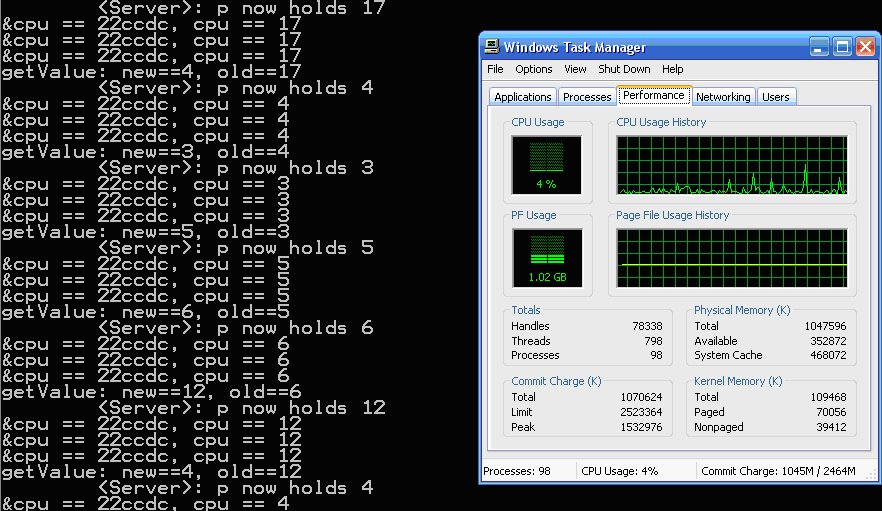}}
\caption{An excerpt from the execution of the code in Fig.~\ref{f:rr1}.}
\label{f:rr3}
\end{figure}

The RR var model does not require any special language: Figure~\ref{f:rr1} is an example in the C language.
The portrayed
program declares two variables: ``cpu'', a reflective integer, which reports the
current level of usage of the local CPU as an integer number between 0 and 100,
and ``tcpTxRate'', a reflective {\em and refractive\/} integer, which reports {\em and sets\/}
the send rate parameter of the TCP layer. The code periodically queries
the CPU usage and, when that reaches a value greater than 90\%, it requests to
change the TCP send rate. Note that the only non standard C construct is
attribute ``Ref\_t'', which specifies that a corresponding declaration is
reflective or refractive or both. Through a translation process
this code is instrumented so as to include the logics required to
interface the cpu and the TCP external devices. Figure~\ref{f:rr3} shows this simple code
in action on our development platform---a Pentium-M laptop running Windows
XP and the Cygwin tools.

We observe that through the RR var model the design complexity is partitioned into two well defined and
separated components: The code to interface external devices is specified in a separate
architectural component,
while the functional code is produced in a familiar way, in this case as a C code
reading and writing integer variables.

The result is a structured model to express tasks such as cross-layered optimization, adaptive or
fault-tolerant computing in an elegant, non intrusive, and cost-effective way. Such model is
characterized by strong separation of design concerns, for the functional strategies are not to be
specified aside with the layer functions; only instrumentation is required, and this can be done once
and for all. This prevents spaghetti-like coding for both the functional and the non-functional aspects,
and translates in enhanced maintainability and efficiency.

A special reflective variable is ``int \_Redundance''. The device attached to \_Redundance
is in this case responsible for providing a trustworthy and timely estimation of the degree of redundancy
matching the current disturbances. This device should be in most cases mission-dependent---for
instance, for electrical substation automation systems, that could be a sensor able to assess
the current electro-magnetic interference. What we consider as a significant property of our
approach is that such dependence is not hidden, but isolated in a custom architectural component. In so doing,
complexity is partitioned but it is also explicitly available to the system designers for inspection 
and maintenance. This can facilitate porting \emph{the service\/} which, as we pointed out
in~\cite{DeDe02b}, is something different from porting \emph{the code\/} of a service.
Failing to do so can bring to awful consequences, as can be seen in well-known accidents
such as the failure of the Ariane 5 flight 501 or, even worse, in the failures of the
Therac-25 linear accelerator~\cite{Lev95}.

In the case at hand---more a proof of concepts than a full fledged product---the device connected to
\_Redundance is just a task that receives the return values of the majority voting
algorithm executed when reading a redundant variable. Such return value is the maximum number of
agreeing replicas in the current voting, which is used to compute the risk of failure,
i.e., variable $r$ in~(\ref{eq:1}).

In the following section we describe how our tool behaves when memory faults are injected.
We shall see that, despite the above na\"ive design choices, our tool already depicts valuable results.

\section{Performance Analysis}\label{s:perfanalysis}
In order to analyze the performance of our system, we have developed a simulator, ``scrambler''.
Our scrambler tool allows to simulate a memory, to protect it with redundant variables, to
inject memory faults (bit flips or ``bursts'' corrupting series of contiguous cells), and to
measure the amount of redundancy actually used. Scrambler interprets a simple scripting language
consisting of the following commands:

\begin{description}
\item[SLEEP $s$], which suspends the execution of the scrambler for $s$ seconds,
\item[SCRAMBLE $n, p$], which repeats $n$ times action ``scramble a pseudo-random 
    memory cell with probability $p$'',
\item[BURST $n, p, l$], which repeats $n$ times action ``scramble $l$ contiguous cells with
    probability $p$'',
\item[END], which terminates the simulation.
\end{description}

The above commands can be used to compose a complex sequence of fault injections. As an example,
the following script, corresponds to the following configuration: no faults for 1 second, then various
disturbances occurring with Gaussian distribution\footnote{%
           The chosen probabilities correspond to Gaussian $f(x)=\exp(-{x^2}/{4})$
           for $x=\pm3,\pm2,\pm1$, and 0.},
then no disturbances again for 5 seconds:

\begin{center}
\begin{verbatim}
SLEEP 1                       // sleep 1 sec
SCRAMBLE 2000, 0.1053992      // scramble 2000 random cells
                              // with probability f(-3)
SCRAMBLE 2000, 0.3678794      // scramble 2000 random cells
                              // with probability f(-2)
BURST 2000, 0.7788008, 10     // execute 2000 bursts of 10 
                              // contiguous cells
                              // with probability f(-1)
SCRAMBLE 2000, 1              // scramble 2000 random cells
BURST 2000, 0.7788008, 10     // execute 2000 bursts of 10
                              // contiguous cells
                              // with probability f(1)
SCRAMBLE 2000, 0.3678794      // scramble 2000 random cells
                              // with probability f(2)
SCRAMBLE 2000, 0.1053992      // scramble 2000 random cells
                              // with probability f(3)
SLEEP 5                       // sleep 5 secs
END                           // stop injecting faults
\end{verbatim}
\end{center}

The idea behind these scripts is to be able to represent executions where a program is
subjected to environmental conditions that vary with time and range from ideal to
heavily disturbed. Scenarios like these ones are common, e.g., in applications servicing
primary substation automation systems~\cite{DSN03} or spaceborne applications~\cite{1981coae.conf...66O}.


In the following we describe a few experiments and the results we obtained with scrambler.

All our experiments have been carried out with an array of 20000 redundant 4-byte cells and an
allocation stride of 20 (that is, replicas of a same logical cell are spaced by 20 physical cells).
In all the reported experiments we run the following script:

\begin{center}
\begin{verbatim}
SLEEP 1
SCRAMBLE 10000, 0.9183156388887342
SCRAMBLE 10000, 0.9183156388887342
SLEEP 3
SCRAMBLE 10000, 0.9183156388887342
SCRAMBLE 10000, 0.9183156388887342
END
\end{verbatim}
\end{center}

Concurrently with the execution of this script, 65 million read accesses were performed in round
robin across the array. The experiments record the number of scrambled cells and the number of
read failures. 

Scrambler makes use of standard C function ``rand'', which depends on an initial seed to generate
each pseudo-random sequence. In the reported experiments the same value has been kept for the seed,
so as to produce exactly the same sequences in each experiment.

\begin{figure}
\centerline{\includegraphics[width=0.4\textwidth]{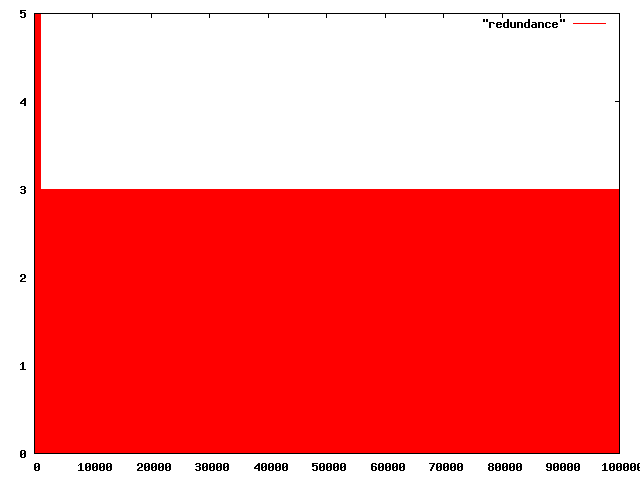}}
\caption{The picture shows how the allocated redundancy varies during the first 100000 cycles 
of Experiment 3. Note how redundancy drops to its minimum after the first 1000 cycles.}
\label{f:ex3}
\end{figure}

\begin{figure}
\centerline{\includegraphics[width=0.6\textwidth]{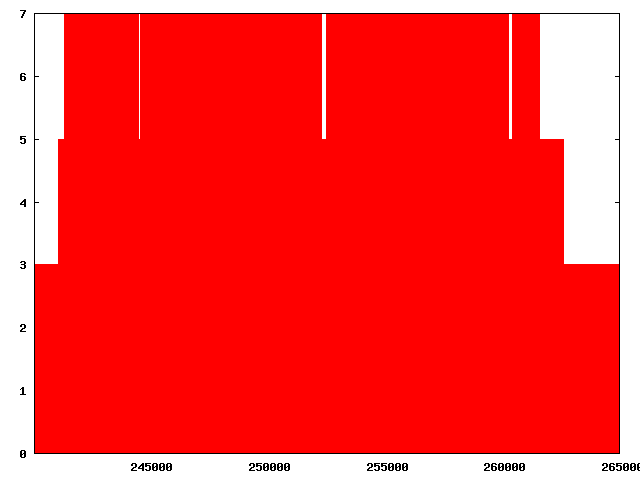}}
\caption{The picture focuses on a section of experiment 3 (namely steps 240150--265000). During
such section, fault injection takes place. Accordingly, the allocated redundancy varies so as to
match the threat being experienced.}
\label{f:ex3:zoom}
\end{figure}

\begin{description}
\item{Experiment 1: Fixed, low redundancy.}
In this first experiment we executed scrambler with fixed (non adaptive) redundancy 3.
Table~\ref{t:ex1} shows the setting of this experiment. The main conclusion we can draw
from this run is that a statically redundant data structures provision in this case fails
111 times: In other words, for 111 times it was not possible to find a majority of replicas
in agreement, and the system reported a read access failure.
The total number of memory accesses is
proportional to $3\times 65000000\times k$, where $k>0$ depends on the
complexity of the redundant read operation.


\begin{table}[t]
\begin{center}
\begin{tabular}{l}
\textbf{\$ scrambler faults.in 3 scrub noadaptive}\\
\texttt{Scrambler::sleep(1)}\\
\texttt{run 1}\\
\texttt{run 50001}\\
\texttt{run 100001}\\
\texttt{run 150001}\\
\texttt{Scrambler::scramble(10000,0.918316)}\\
\texttt{Scrambler::scramble(10000,0.918316)}\\
\texttt{Scrambler::sleep(3)}\\
\texttt{run 200001}\\
\texttt{run 250001}\\
\emph{\ldots lines omitted\ldots}\\
\texttt{run 650001}\\
\texttt{Scrambler::scramble(10000,0.918316)}\\
\texttt{Scrambler::scramble(10000,0.918316)}\\
\texttt{Scrambler::END}\\
\texttt{run 700001}\\
\texttt{run 750001}\\
\emph{\ldots lines omitted\ldots}\\
\texttt{run 65000001}\\
\texttt{36734 scrambled cells, 189 failures, redundance == 3}\\
\texttt{redundance 3: 65000001 runs}\\
\texttt{redundance 5: 0 runs}\\
\texttt{redundance 7: 0 runs}\\
\texttt{redundance 9: 0 runs}\\
\texttt{redundance 11: 0 runs}
\end{tabular}
\end{center}
\caption{Experiment 1: Scrambler executes the script in file ``faults.in''. Parameters 
set redundancy to 3, select memory scrubbing to repair corrupted data when possible, and
keep redundancy fixed.}\label{t:ex1}
\end{table}

\item{Experiment 2: Fixed, higher redundancy.}
Also experiment 2 has fixed redundancy, this time equal to 5.
Table~\ref{t:ex2} shows the setting of this new experiment. 
Main conclusion is that the higher redundancy is enough to guarantee data integrity in this
case: No read access failures are experienced. The total number of memory accesses is
proportional to $5\times 65000000\times k$.

\begin{table}
\begin{center}
\begin{tabular}{l}
\textbf{\$ scrambler faults.in 5 scrub noadaptive}\\
\texttt{Scrambler::sleep(1)}\\
\texttt{run 1}\\
\texttt{run 50001}\\
\texttt{run 100001}\\
\texttt{Scrambler::scramble(10000,0.918316)}\\
\texttt{Scrambler::scramble(10000,0.918316)}\\
\texttt{Scrambler::sleep(3)}\\
\texttt{run 150001}\\
\emph{\ldots lines omitted\ldots}\\
\texttt{run 500001}\\
\texttt{Scrambler::scramble(10000,0.918316)}\\
\texttt{Scrambler::scramble(10000,0.918316)}\\
\texttt{Scrambler::END}\\
\texttt{run 550001}\\
\emph{\ldots lines omitted\ldots}\\
\texttt{run 65000001}\\
\texttt{36734 scrambled cells, 0 failures, redundance == 5}\\
\texttt{redundance 3: 0 runs}\\
\texttt{redundance 5: 65000001 runs}\\
\texttt{redundance 7: 0 runs}\\
\texttt{redundance 9: 0 runs}\\
\texttt{redundance 11: 0 runs}
\end{tabular}
\end{center}
\caption{Experiment 2: Scrambler executes the same script as before; only, redundancy is now set 
to 5. No failures are observed.}\label{t:ex2}
\end{table}

\item{Experiment 3: Adaptive redundancy.}
In this last experiment we enable adaptive redundancy which we initially set to 5.
Table~\ref{t:ex3} shows the resulting setting. Most worth noting is the fact that
also in this case no read access failures show up, but the actual amount of redundancy
required to reach this result is much lower.
Consequently, also the total number of memory accesses,
proportional to $(3\times 64953188 + 5\times 5631 + 7\times 26534 + 9\times 14648)\times k$,
is considerably lower.
Figure~\ref{f:ex3} shows how redundancy varies during the first 100000 read cycles.
During this time frame no fault injection takes place. This is captured by our
adaptation strategy, which decreases redundancy to 3 after 1000 cycles.
Figure~\ref{f:ex3:zoom} depicts an interval where several fault injections do take place.
These events are detected and trigger several adaptations.
\end{description}

\begin{table}[t]
\begin{center}
\begin{tabular}{l}
\textbf{\$ scrambler faults.in 5 scrub adaptive}\\
\texttt{run 1}\\
\texttt{Scrambler::sleep(1)}\\
\texttt{run 50001}\\
\texttt{run 100001}\\
\texttt{run 150001}\\
\texttt{run 200001}\\
\texttt{Scrambler::scramble(10000,0.918316)}\\
\texttt{Scrambler::scramble(10000,0.918316)}\\
\texttt{Scrambler::sleep(3)}\\
\texttt{run 250001}\\
\emph{\ldots lines omitted\ldots}\\
\texttt{run 600001}\\
\texttt{Scrambler::scramble(10000,0.918316)}\\
\texttt{Scrambler::scramble(10000,0.918316)}\\
\texttt{Scrambler::END}\\
\texttt{run 650001}\\
\emph{\ldots lines omitted\ldots}\\
\texttt{run 65000001}\\
\texttt{36734 scrambled cells, 0 failures, redundance == 3}\\
\texttt{redundance 3: 64953188 runs}\\
\texttt{redundance 5: 5631 runs}\\
\texttt{redundance 7: 26534 runs}\\
\texttt{redundance 9: 14648 runs}\\
\texttt{redundance 11: 0 runs}
\end{tabular}
\end{center}
\caption{Experiment 3: Scrambler executes the same script as before; only, redundancy is now adaptive.
No failures are observed, but the employed redundancy is mostly of degree 3.}\label{t:ex3}
\end{table}

\section{Conclusions}\label{s:end}
As Einstein said, we should ``make everything as simple as possible, but not simpler''.
Likewise, hiding complexity is good, but hiding too much can lead to disasters---history of computing is
paved with noteworthy examples.
One of the main conclusions of this paper is that software-intensive systems 
should be built with architectural and/or structuring
techniques able to decompose the software complexity, but \emph{without hiding the basic hypotheses and
assumptions about their target execution environment and the expected fault- and system models}. A
judicious assessment of what should be made transparent and what should stay translucent is necessary.
A major lesson learned from the experiences reported here is that one of the aspects 
where transparency is not an issue is resiliency. In particular,
the fault model should be explicit, or severe mistakes can result.
Also, as environments do change, e.g. because of external events, or because assets dissipate, or because the
service is mobile, there is no static allocation of resources that can accommodate for any possible
scenario: A highly redundant system will withstand no more faults than those considered at design time,
and will allocate a large amount of resources even when no faults are threatening the service.

A core idea in this paper is that software-intensive systems must be structured so as to allow a transparent
``tuning" to the constantly changing environmental conditions and technological dependencies. A separated
layer must manage monitoring and adaptation, either autonomously or in concert with other system
management layers (e.g. the OS or the middleware). Though separated, the intelligence to achieve
monitoring or adaptation must not be hidden, but encapsulated in well-defined architectural components.
By doing so, it becomes possible to review, verify and maintain that intelligence even when the
hypotheses that were originally valid at design time would change. 

This paper also introduced and discussed a tool that follows these ideas and provides 
a simple, well-defined, and effective system structure for the expression 
of data integrity in the application software. Such tool has a public side, where the functional
service is specified by the user in a familiar form--that of common programming languages such as C---and
a private side, separated but not hidden, where the adaptation logics is defined. Such private side
is the ideal ``location'' to specify fault model tracking, failure semantics adaptation, resource
(re-)optimization, and other important non-functional design goals. An important property to
achieve this is reflection, which is useful to define a homogeneous framework where etherogeneous
system properties can be monitored and reasoned upon.

We believe that our tool represents one of the practical 
``ways to enhance the ability of
organizations to create processes that are robust yet flexible, to monitor and revise risk models, and
to use resources proactively in the face of disruptions or ongoing production and 
economic pressures''~\cite{HWL06,Sim07}, which are sought by the novel discipline of 
resilience engineering. Software design urgently requires tools, methodologies, and
architectures compliant to such model, which we refer to here as a ``new software development.''


\bibliographystyle{latex8}


\end{document}